\documentclass[aps,prl,twocolumn,showpacs,superscriptaddress]{revtex4-1}

\usepackage{amsfonts}
\usepackage{amsmath}
\usepackage{amssymb}
\usepackage{graphicx}

\begin{document}

%%%%%%%%%%%%%%%%%%%%%%%%%%%% TITLE

\title{Systematics of the temperature-dependent inter-plane resistivity in Ba(Fe$_{1-x}$T$_x$)$_2$As$_2$ with T= Rh, Ni, and Pd}

%%%%%%%%%%%%%%%%%%%%%%%%%%%% AUTHORS

\author{M.~A.~Tanatar}
\email[Corresponding author: ]{tanatar@ameslab.gov}
\affiliation{Ames Laboratory US DOE, Ames, Iowa 50011, USA}

\author{N.~Ni}

\affiliation{Ames Laboratory US DOE, Ames, Iowa 50011, USA}
\affiliation{Department of Physics and Astronomy, Iowa State University, Ames, Iowa 50011, USA }

\author{A.~Thaler}

\affiliation{Ames Laboratory US DOE, Ames, Iowa 50011, USA}
\affiliation{Department of Physics and Astronomy, Iowa State University, Ames, Iowa 50011, USA }

\author{S.~L.~Bud'ko}
\affiliation{Ames Laboratory US DOE, Ames, Iowa 50011, USA}
\affiliation{Department of Physics and Astronomy, Iowa State University, Ames, Iowa 50011, USA }

\author{P.~C.~Canfield}
\affiliation{Ames Laboratory US DOE, Ames, Iowa 50011, USA}
\affiliation{Department of Physics and Astronomy, Iowa State University, Ames, Iowa 50011, USA }

\author{R.~Prozorov}
%\email[Corresponding author: ]{prozorov@ameslab.gov}
\affiliation{Ames Laboratory US DOE, Ames, Iowa 50011, USA}
\affiliation{Department of Physics and Astronomy, Iowa State University, Ames, Iowa 50011, USA }

\date{11 May 2011}

%%%%%%%%%%%%%%%%%%%%%%%%%%%% ABSTRACT

\begin{abstract}

Temperature-dependent inter-plane resistivity, $\rho _c(T)$,  was measured systematically as a function of transition metal substitution in the iron-arsenide superconductors Ba(Fe$_{1-x}$T$_x$)$_2$As$_2$, $T$= Ni, Pd, Rh. The data are compared with the behavior found in Ba(Fe$_{1-x}$Co$_x$)$_2$As$_2$, revealing resistive signatures of pseudogap.
In all compounds we find resistivity crossover at a characteristic pseudogap temperature $T^*$ from non-metallic to metallic temperature dependence on cooling. Suppression of $T^*$ proceeds very similar in cases of Ni and Pd doping and much faster than in similar cases of Co and Rh doping. In cases of Co and Rh doping an additional minimum in the temperature-dependent $\rho _c$ emerges for high dopings, when superconductivity is completely suppressed. These features are consistent with the existence of a charge gap covering part of the Fermi surface. The part of the Fermi surface affected by this gap is notably larger for Ni and Pd doped compositions than in Co and Rh doped compounds.
\end{abstract}


% 74.70.Xa	Pnictides and chalcogenides
% 72.15.-v	Electronic conduction in metals and alloys
% 74.62.-c	Transition temperature variations, phase diagrams

\maketitle

%%%%%%%%%%%%%%%%%%%%%%%%%%%% INTRODUCTION

\section{Introduction}
%Oxypnictide superconductors, discovery, properties

Pseudogap or partial gap in the electronic structure, affecting some regions of the Fermi surface while leaving other unaffected, is one of the key signatures of the underdoped cuprates \cite{TS}. It is revealed through anomalous behaviors of the temperature-dependent resistivity, magnetization, NMR Knight shift and relaxation rate, as well as in spectroscopic data \cite{TS}. Pseudogap shows the same
$k$-space distribution as the superconducting gap \cite{Ronning,Kaminski} and is universally observed in hole- and electron- doped cuprates in the underdoped regime.

Features consistent with pseudogap are also clearly found in hole doped FeAs-based materials (see \cite{ishida} for a review).  Because the parent compounds of iron pnictides are metals, the pseudogap here is believed to arise from nesting instability \cite{ishida}.

In Ba(Fe$_{1-x}$T$_x$)$_2$As$_2$ (BaT122 in the following)substitution of the transition metals into Fe position leads to electron doping. NMR studies suggest the existence of a pseudogap in BaCo122 over the broad doping range including full domain of superconductivity, from magnetically ordered parent compound to non-superconducting metal. Existence of pseudogap leads to a temperature-dependent Knight shift $K$, well described by a formula $K=A+B\times exp (-T_{PG}/T)$, where the first term describes contribution of the metallic portion of the Fermi surface and the second activated term allows determination of the $T_{PG} \equiv \Delta_{PG}/k_B$ as 560~K $\pm$150~K at optimal doping \cite{BaCoNMRPG1,BaCoNMRPG2,BaCoNMRPG3}. At temperatures $T < T ^* \ll T_{PG}$ this leads to temperature independent Knight shift and a crossover to metallic temperature dependence in the inter-plane resistivity  \cite{pseudogapCo}. No discernible features are observed in the in-plane resistivity \cite{NiNiCo,NDL}, which suggests that the areas of the Fermi surface affected by pseudo-gap are rather small and belong to the most warped parts of the Fermi surface, contributing mostly to inter-plane transport.

In this article we report a systematic study of the evolution of the inter-plane resistivity with doping by other transition metals inducing superconductivity in Ba122, $T$= Rh, Ni, Pd. We show that similar anomalies are observed in the temperature dependent inter-plane resistivity for all types of substitution, with the characteristic temperature of resistive crossover being suppressed with doping. The rate of $T^*$ suppression with $x$ is however notably higher in Ni and Pd doped compositions, even with correction for a difference in number of added electrons. The doping-dependence of the pseudogap feature suggests that it represents an independent energy scale in the problem, different from that of structural/magnetic transition and superconductivity.

%%%%Experimental
\section{Experimental}

Single crystals of BaFe$_2$As$_2$ doped with Ni, Pd and Rh were grown as described in detail in previous communications \cite{NiNiNi,NiCuscaling,NiPdRh}. Crystals were thick platelets with large faces corresponding to the tetragonal (001) plane.  The actual content of transition metals, $x$, was determined with wavelength dispersive spectroscopy (WDS) electron probe microanalysis and is used in the following.

We used two-probe resistivity measurements, as justified by ultralow contact resistance of Sn-soldered contacts \cite{SUST}. The details of sample preparation, sample screening and selection are identical to those used in our studies of $c$-axis resistivity in Co-doped material \cite{anisotropy,anisotropypure,pseudogapCo}. The absolute values of the inter-plane resistivity at room temperature for most compositions stays in the range 1 to 1.5 m$\Omega$cm, with doping it decreases to approximately 0.5 m$\Omega$cm. For several $x$ compositions we were not able to find crystals with resistivity values lower than 2~  m$\Omega$cm, despite the facts that (1) the evolution of the temperature-dependent resistivity for these samples followed the general trend, (2) close in $x$ compositions show usual resistivity values. This limits the accuracy of the absolute $\rho _c$ value determination by approximately a factor of two.

%%%Results
\section{Results}

\begin{figure}
		\includegraphics[width=4cm]{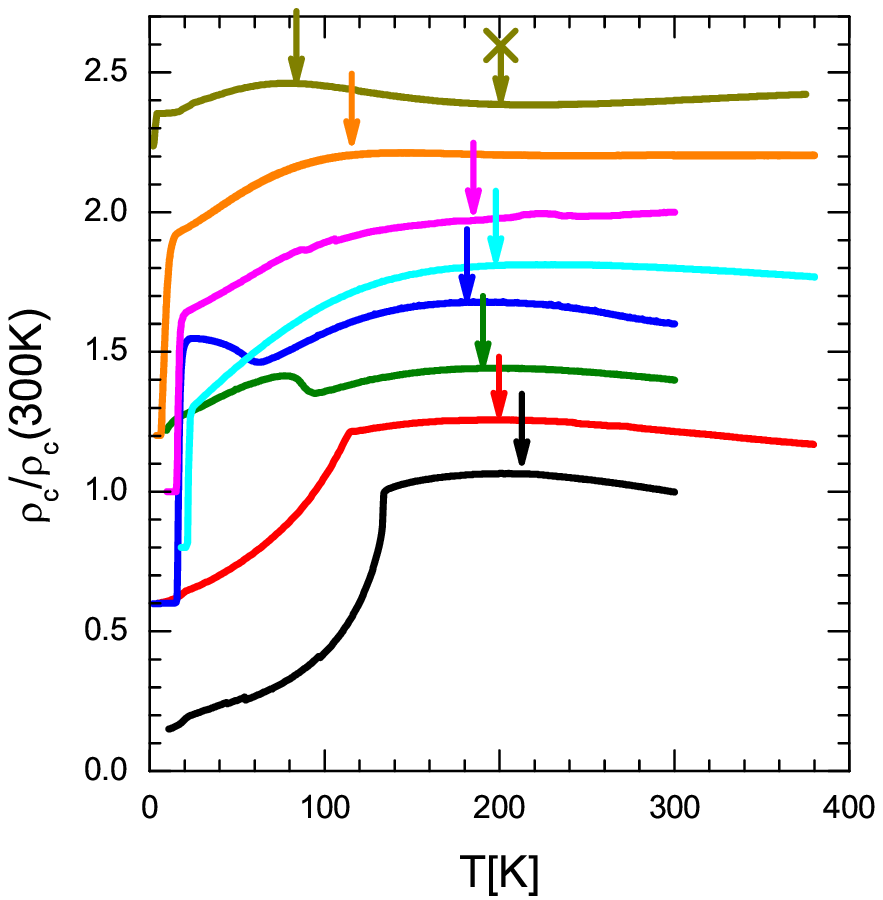}
		\includegraphics[width=4cm]{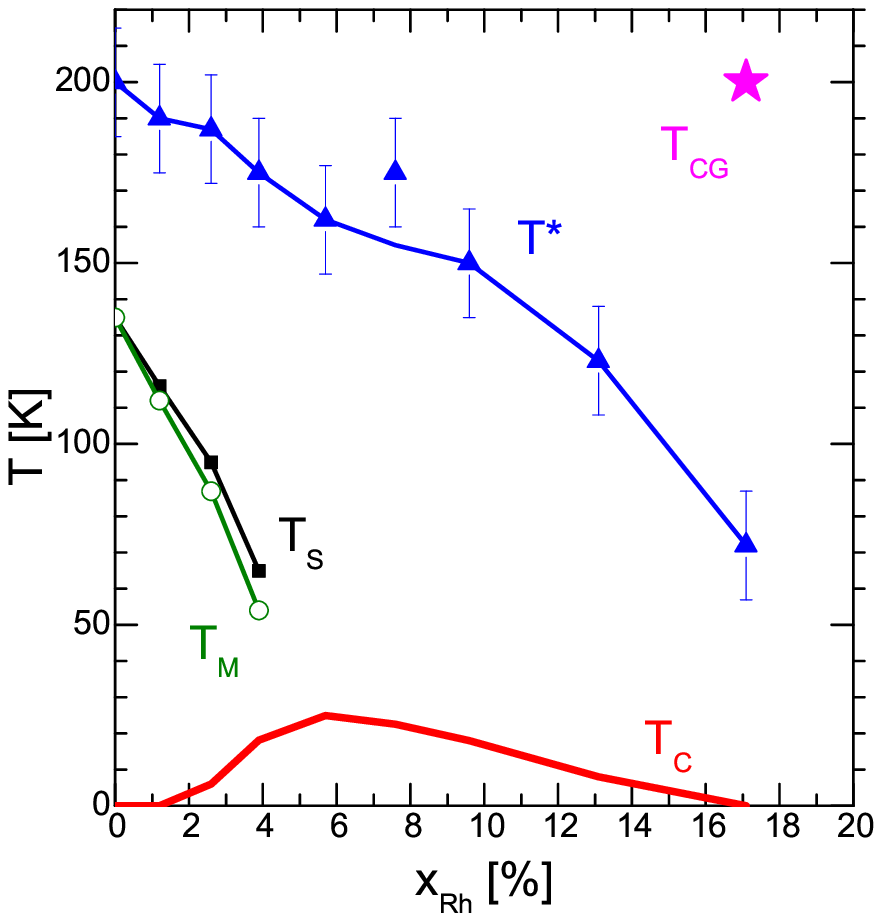}
\caption{Temperature dependence of the inter-plane resistivity, $\rho_c $, normalized by its value at room temperature $\rho_c (300K)$, for samples of Ba(Fe$_{1-x}$Rh$_x$)$_2$As$_2$ with $x \le 0.171$ (slightly above the concentration boundary for the superconducting dome). Lines are offset, from bottom to top, $x_{Rh}$= 0, 0.012, 0.026, 0.039, 0.076, 0.096, 0.131 and 0.171 (top panel).  Arrows show a position of the resistivity maximum, $T^*$, cross-arrow shows a position of the resistivity minimum $T_{CG}$. Bottom panel shows the $T-x_{Rh}$ phase diagram, in which lines of structural $T_S$, magnetic $T_M$ and superconducting $T_c$ states are determined from in-plane resistivity and magnetization measurements, see Ref.~\onlinecite{NiPdRh}.
}
\label{Rh}
\end{figure}

In the top panel of Fig.~\ref{Rh} we plot inter-plane resistivity of Ba(Fe$_{1-x}$Rh$_x$)$_2$As$_2$, using normalized scale $\rho _c/ \rho _c(300K)$. To avoid overlapping, the curves are offset progressively upwards for higher dopings. The data for parent compound are reproduced from Ref.~\onlinecite{anisotropypure}. Several features of the temperature dependence are essentially the same as observed in compositions with Co doping \cite{pseudogapCo}. For low doping $x_{Rh}$=0.012, $\rho _c (T)$ shows resistivity increase on cooling for $T > T^*$ (marked with an arrow), which is the pseudogap feature in all compositions studied. On reaching a temperature of structural/magnetic transition $T_S$, resistivity dives down, and decreases monotonically all the way to base temperature, showing some signatures of filamentary superconductivity at about 20~K due to strain \cite{Saha}. This decrease of resistivity below $T_S$ is similarly observed in lightly Co doped composition $x_{Co}$= 0.012 \cite{pseudogapCo}. For higher dopings $x_{Rh}$=0.026 and $x_{Rh}$=0.39 resistivity shows increase below $T_S$ and drop to zero below superconducting $T_c$, in complete accord to the behavior found in Co-doped compositions.  Finally, for compositions in which structural/magnetic transition is completely suppressed, $x > \sim 0.07$, position of the resistivity maximum shifts down in temperature and for an ultimate doping $x_ {Rh}$= 0.171, when superconductivity is completely suppressed, the resistivity shows shallow minimum, marked with cross arrow in Fig.~\ref{Rh}. The fact of the appearance of the minimum at high dopings and even the doping value at which it appears are very similar to those found in Co-doped compounds. We summarize our observations in the temperature-doping, $T$-$x$, phase diagram in bottom panel of Fig.~\ref{Rh}.

%%%%%%%%%%%%%%Fig rhoc(T) vs Co concentration, phase diagram, roa (T) vs Co

\begin{figure}
		\includegraphics[width=8cm]{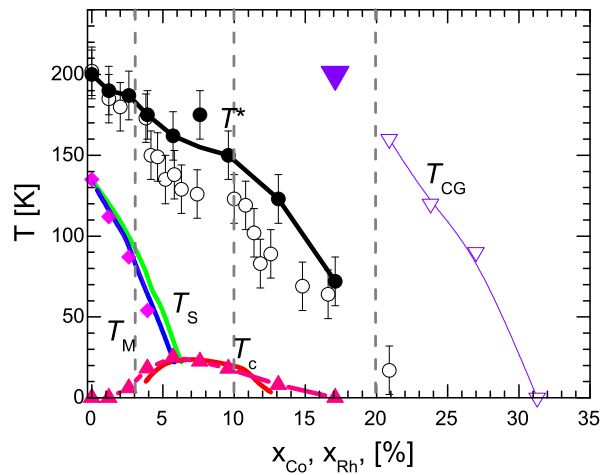}
	\caption{Temperature- concentration phase diagrams of the pseudogap features as revealed from the temperature dependent inter-plane resistivity, $\rho_c $, in $T$=Rh (solid symbols, this study) and $T$=Co (open symbols, Ref.~\onlinecite{pseudogapCo}) doped Ba(Fe$_{1-x}$T$_x$)$_2$As$_2$. Pseudogap features for two types of doping overlap within error bars, similar to lines of structural $T_S$, magnetic $T_M$ and superconducting $T_c$ transitions, Ref.~\onlinecite{NiPdRh}. Vertical lines separate composition ranges in which Fermi surface topology changes in $T$=Co \cite{Lifshits1,Lifshits2,Lifshits3}.
}
	\label{RhCo}
\end{figure}

In Fig.~\ref{RhCo} we compare explicitly the phase diagrams of the pseudogap features in $c$-axis resistivity of Co and Rh doped compounds. Within rather big error bars due to crossover character of features, the diagrams are overlapping. This fact is remarkable, since despite electronic equivalence of Co and Rh doping, the $c$-axis lattice parameters are different in two materials \cite{NiPdRh}, and thus it would be natural that this difference should affect characteristic energy scales of the electronic overlap in the interplane direction.

%%%%Ni and Pd doping

\begin{figure}
		\includegraphics[width=8cm]{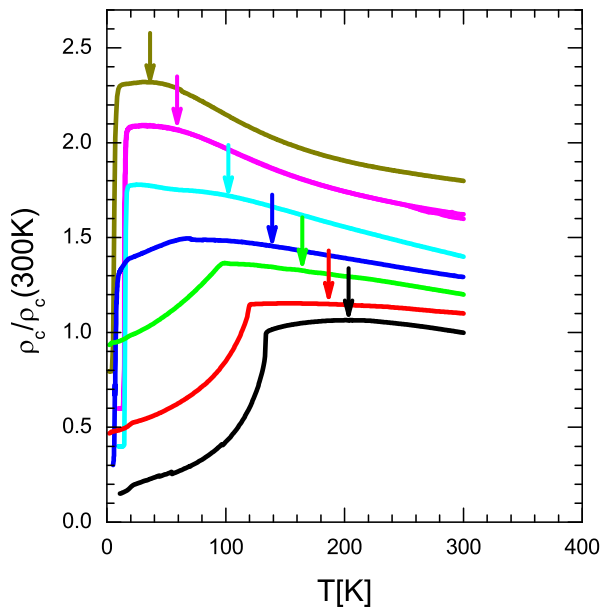}
	\caption{Temperature dependence of the inter-plane resistivity, $\rho_c $, normalized by its value at room temperature $\rho_c (300K)$, for samples of Ba(Fe$_{1-x}$Ni$_x$)$_2$As$_2$ with $x \le 0.072$ (slightly above the concentration boundary for the superconducting dome). Lines are offset, from bottom to top, $x_{Ni}$= 0, 0.0067, 0.016, 0.024, 0.032, 0.054, and 0.072 (top panel).  Arrows show a position of the resistivity maximum, $T^*$, used to plot the phase diagram, see Fig.~\ref{NiPdphaseD}.
}
	\label{Ni}
\end{figure}

\begin{figure}
		\includegraphics[width=8cm]{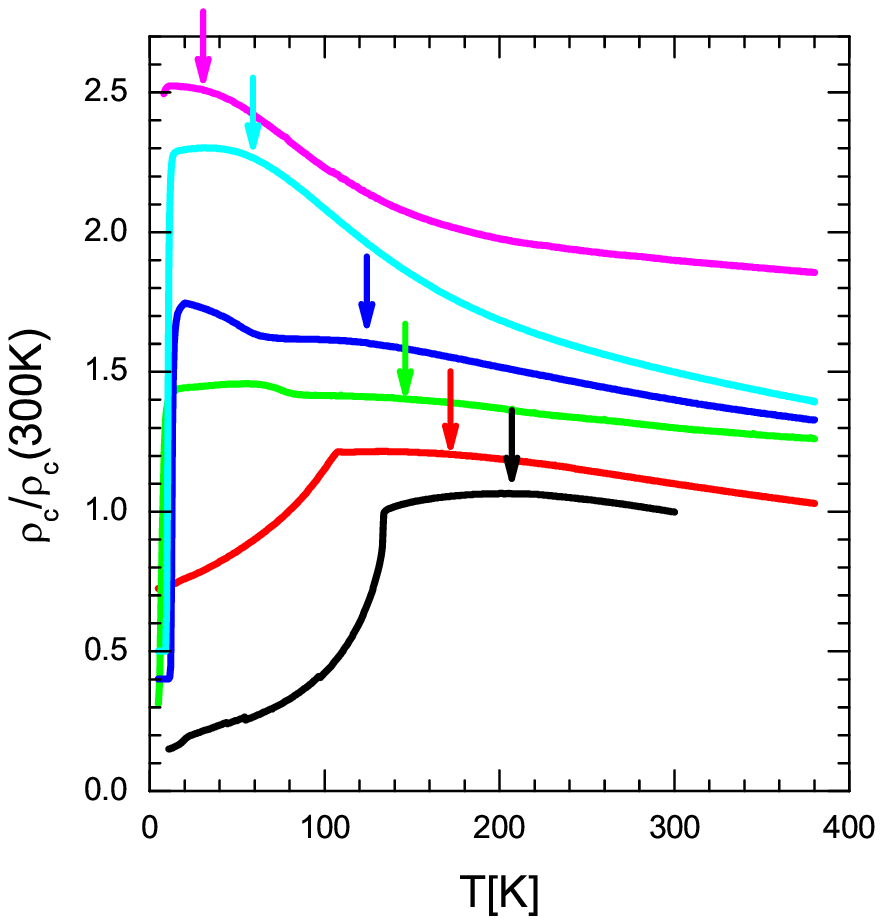}
\	\caption{Temperature dependence of the inter-plane resistivity, $\rho_c $, normalized by its value at room temperature $\rho_c (300K)$, for samples of Ba(Fe$_{1-x}$Pd$_x$)$_2$As$_2$ with $x \le 0.077$ (slightly above the concentration boundary for the superconducting dome). Lines are offset, from bottom to top, $x_{Pd}$= 0, 0.012, 0.021, 0.027, 0.030, 0.053, and 0.077 (top panel).  Arrows show a position of the resistivity maximum, $T^*$, used to plot the phase diagram, see Fig.~\ref{NiPdphaseD}.
}
	\label{Pd}
\end{figure}

In Fig.~\ref{Ni} and Fig.~\ref{Pd} we show doping-evolution of the temperature-dependent interplane resistivity in samples doped with transition metals of group 10 of Mendeleev periodic table, 3$d$ $T$=Ni and 4$d$ $T$=Pd. These atoms donate two electrons on substituting Fe, and thus substitution level required to induce superconductivity is two times lower than in cases of Co and Rh doping.
Bottom panels of the Fig.~\ref{Ni} and Fig.~\ref{Pd}, show doping phase diagram of the pseudogap features in interplane resistivity.

The suppression of characteristic temperature of the resistivity maximum, $T^*$, is much more rapid for cases of Ni and Pd doping, and $T^* (x)$ line in the phase diagram suggests critical concentration very close to the edge of the superconducting dome. Moreover, for highest doping levels resistivity monotonically decreases with temperature and does not reveal a minimum at $T_{CG}$ as in cases of Co and Rh doping. On the other hand, similar to the case of Co and Rh, the phase diagrams of Ni and Pd dopings coincide within error bars, see top panel of Fig.~\ref{NiPdRh}.

\begin{figure}
		\includegraphics[width=8cm]{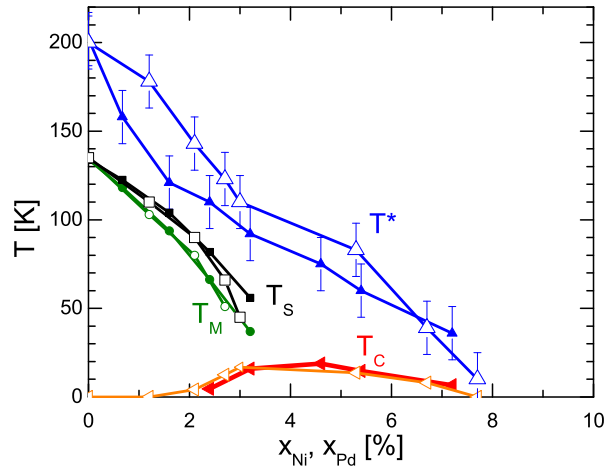}
\	\caption{ Comparison of the temperature- concentration phase diagrams of the pseudogap features in $T$=Ni (solid symbols) and $T$=Pd (open symbols) doped Ba(Fe$_{1-x}$T$_x$)$_2$As$_2$ as determined from inter-plane resistivity measurements. Lines of structural $T_S$, magnetic $T_M$ and superconducting $T_c$ states are determined from in-plane resistivity and magnetization measurements, see Ref.~\onlinecite{NiPdRh}. All characteristic features for two types of doping overlap within error bars.
}
	\label{NiPdphaseD}
\end{figure}

\section{Discussion}

\subsection{Scaling relations for various dopants}

\begin{figure}
		\includegraphics[width=4cm]{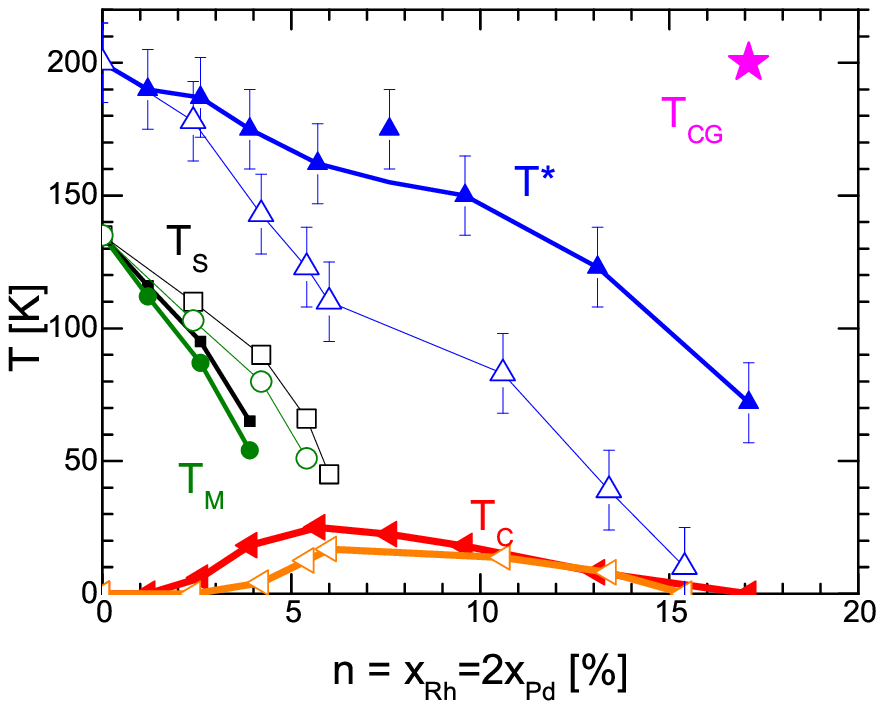}
		\includegraphics[width=4cm]{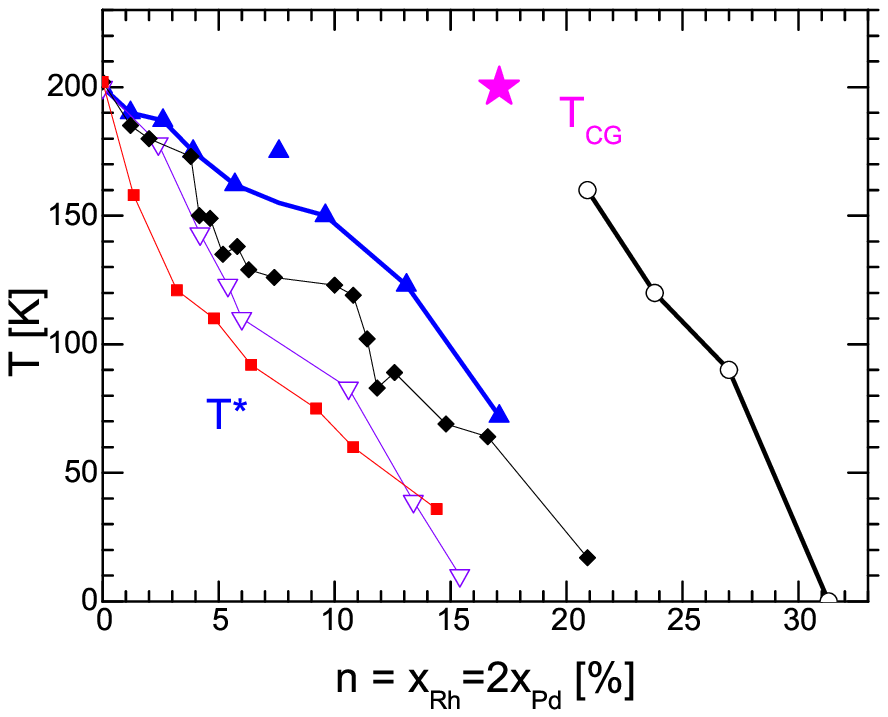}
	\caption{Top panel: comparison of the doping phase diagrams for Rh (solid symbols) and Pd (open symbols) doping, plotted vs concentration of added electrons defined as $n= x_{Rh}=2x_{Pd}$. Bottom panel shows phase diagram of the pseudogap features for all studied transition metals. Resistivity maximum, $T^*$, for Rh (blue solid up-triangles), Co (black solid diamonds), Pd (open down-triangles) and Ni (red solid squares) and resistivity minimum, $T_{CG}$, for Rh (solid pink star) and Co (open black circles).
}
	\label{NiPdRh}
\end{figure}

In systematic study of doping phase diagrams for a variety of transition metal dopants: Co, Rh, Ni, Pd, Cu and Cu+Co, it was found that the superconducting transition temperature $T_c$ scales with the number of doped electrons, $n=x_{Co}=x_{Rh}=2x_{Ni}=2x_{Pd}=3x_{Cu}$, while the structural/magnetic transition temperatures scale with number of dopant atoms  $x=x_{Co}=x_{Rh}=x_{Ni}=x_{Pd}=x_{Cu}$.
It is therefore interesting if pseudogap features follow either of these scaling relations. While the break-up of $T^*(x)$ scaling is obvious from the comparison of Fig.~\ref{RhCo} and Figs.~\ref{Ni},~\ref{Pd}, in bottom panel of Fig.~\ref{NiPdRh} we compare the phase diagrams of Rh and Pd doped BaFe$_2$As$_2$ vs the number of extra electrons $n$. It is clear that the pseudogap temperature does not scale with $n$, contrary to the supercoducting transition temperature. Because of relatively big error bar in determination of the pseudogap features, in the bottom panel of Fig.~\ref{NiPdRh} we compare pseudogap features for all four compounds studied. The difference between the two columns is still clearly resolved, supporting the lack of the scaling of pseudogap features with $n$. The difference in doping dependence of pseudogap features pushes us to recognize pseugogap as an independent characteristic scale in the phase diagram of Ba122 materials.

\subsection{Relation to Fermi surface topology}

The evolution of the Fermi surface topology in BaCo122 with doping was studied using ARPES and thermopower measurements \cite{Lifshits1,Lifshits2,Lifshits3}. A sequence of three Lifshits transitions was found, with concentration boundaries at $x_1$=0.3, $x_{2 \Gamma}$=0.11 and $x_{2 Z}$=0.195 \cite{Lifshits1,Lifshits2,Lifshits3}. These are shown with grey lines in Fig.~\ref{RhCo}. Unexpectedly, doping evolution of the pseudogap temperature $T^*$ shows little correlation with characteristic features in the Fermi surface evolution. In particular, hole pocket near $\Gamma$ point in the Brillouin zone changes shape from cylindrical to ellipsoidal at $x_{2 \Gamma}$, but merely any feature can be noticed in $T^* (x)$, see Fig.~\ref{RhCo}. Considering low anisotropy of electrical resistivity of the compounds, it is not clear which cylinder of the Fermi surface is responsible for carrier activation, however, pseudogap affects most strongly the most warped portions.

Comparison of the temperature-dependent inter-plane resistivities for heavily overdoped Rh $x_{Rh}$=0.171 and Pd $x_{Pd}$=0.077 shows an interesting difference. The resistivity monotonically decreases with heating for $x_{Pd}$=0.077, top curve in Fig.~\ref{Pd}, decreasing from low temperatures to room temperature by a factor of 2. Very similar magnitude of decrease is observed in sample with $x_{Ni}$=0.072, top curve Fig.~\ref{Ni}. The magnitude of decrease in $x_{Rh}$=0.171, top curve in Fig.~\ref{Rh}, is merely 10\%, and the curve shows metallic resistivity increase on heating above $\sim$200~K. This difference suggests that the intact metallic part of the Fermi surface, contributing to the interplane transport, is notably smaller for Ni and Pd doping, than in cases of Co and Rh doping. It is interesting if this difference can be found in NMR studies as well.

\section{Conclusion}

In conclusion, our systematic study puts constraints on the possible origin of the anomalous resistivity behavior in transition metal doped BaFe$_2$As$_2$. Characteristic temperatures of the pseudogap do not follow inter-plane distance, in the samples of the same column of Mendeleev table, as would be naturally expected for variation of the inter-plane transfer integrals. They neither follow the electron count as would be expected if they were in relation with the Fermi surface volume/topology (in contrast with the superconducting $T_c$ which scales with $n$), nor doping evolution of the structural/ magnetic transitions.  This pushes us to recognize pseudogap as yet another energy scale in the fascinating complexity of the iron pnictide superconductors.
Its origin remains as obscure at the moment as the origin of mysterious pseudogap phase in the cuprates.

%%%%%%%%%%%%%%%%%%%%%%%%%%%% ACKNOWLEDGMENTS
\section{Acknowledgements}

We thank A. Kreyssig for useful discussions. Work at the Ames Laboratory was supported by the Department of Energy-Basic Energy Sciences under Contract No. DE-AC02-07CH11358. SLB was supported in part by the State of Iowa through the Iowa State University.

%%%%%%%%%%%%%%%%%%%%%%%%%%%% BIBLIOGRAPHY

\end{document}